\documentclass[twocolumn,showpacs,superscriptaddress,floatfix]{revtex4}
\usepackage{graphicx}
\usepackage{dcolumn}
\usepackage{bm}
\usepackage{subfigure}
\begin{document}
\title{Polarization transitions in interacting ring 1D arrays}
\author{Bahman Roostaei}
\affiliation{Dept. of Phys. and Ast., University of Oklahoma, Norman,OK 73019}
\author{Kieran J. Mullen,}
\affiliation{Dept. of Phys. and Ast., University of Oklahoma, Norman,OK 73019}
\date{\today}
\begin{abstract}
Periodic nanostructures can display the dynamics of arrays of atoms 
while enabling the tuning of interactions in
ways not normally possible in Nature.  
We examine one dimensional arrays 
of a ``synthetic atom,'' a one dimensional ring with a nearest neighbor 
Coulomb interaction.  We consider the
classical limit first, finding that the singly charged rings possess
antiferroelectric order at low temperatures when the charge is discrete,
but that they do not order when the charge is treated as a continuous
classical fluid.  In the quantum limit Monte Carlo simulation suggests that 
the system undergoes a quantum phase
transition as the interaction strength is increased.  This is supported by
mapping the system to the 1D transverse field Ising model.  Finally we
examine the effect of magnetic fields.   We find that a magnetic field can
alter the electrostatic phase transition producing a ferroelectric groundstate, 
solely through its effect of shifting the eigenenergies of the quantum problem.

\end{abstract}
\maketitle
\section{Introduction}
Fabrication of micro- and nano-sized structures such as quantum
dots, wires and rings has made it possible for
physicists to examine new ideas in electronic devices.
These small size devices act as artificial atoms with spectra and
shell structures similar to those of real atoms\cite{nano}. 
However it is possible to control 
the properties
of these synthetic atoms in a way that is impossible with real ones.
For example, 
a regular
atomic orbital is three dimensional and one has limited control over the
electronic wavefunction. In contrast, by controlling the shape
of a quandum dot we can distort the wavefunction, controlling its
polarizability and its interaction with adjacent dots. 
While the properties
of periodic arrays of atoms
 are well understood in solid state physics, 
we have new playing field -- periodic arrays of nanostructures --
in which we have an unprecedented control of the ``atomic states.''

The focus of this paper is on the periodic one dimenstional (1D) 
arrays  of nano-rings.   We chose this system for two reasons.  First,
quantum rings display interesting phenomena (e.g. persistent
currents\cite{PersistentTheory1,PersistentTheory2,PersistentExpt1,PersistentExpt2}. 
)
which are not found in dots.  
The basic difference between
the ring geometry and a quantum dot is the excluded middle which
confines the electron in a ring to a narrow spatially periodic
channel. This compact, periodic geometry can allow dynamics not 
found in other systems.\cite{compact}
Second, it has been become possible to create extremely small rings.
These arrays of nano-rings can be fabricated either by dry etching
\cite{MattRings} or by
using MBE techniques to foster self assembled InGaAs/GaAs rings.
The size of these nano-rings is $\sim$30~nm for outer radius and
$\sim$10~nm for inner radius for self-assembled InGaAs/GaAs rings
\cite{GaAsRings}.    Such techniques not only produce extremely small
rings, they also make it easy to make {\it periodic} arrays of small rings.

In this paper we consider an
ideal array of 1D rings at zero temperature, each carrying a
single charge.  The rings are sufficiently close together that there
is a Coulomb interaction between the electrons, but separated enough
so that the tunneling between rings can be neglected,\cite{limits}
as discussed in section \ref{sec:model}.  In section \ref{sec:classical}
we consider the classical case in section and see how
different assumptions allow for symmetric or symmetry breaking
ground states.   Section \ref{sec:quantum} contains the main results
of this work, where we show that there is a {\it quantum phase
transition} in a 1D array of rings for $B=0$.  We show this both through Monte
Carlo  simulation of $\rm 1+1$D classical statistical respresentation
of the problem as well as mapping it on to the 1D transverse field
Ising model.  The polarization pattern is antiferroelectric. 
In section \ref{sec:Bfield} we examine how this transition is affected
by magnetic fields.  
We
conclude in section \ref{sec:Conclusions} 
by summarizing results and discussing possible applications.

\begin{figure}
  \includegraphics[width=1.7in]{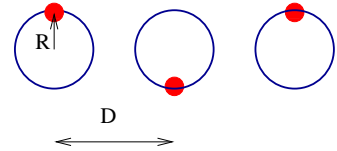}\\
  \caption{A schematic picture of the groundstate of
classical point electrons for 1D array of rings. The ring radius is R and the separation is D. The 1D ordering
is antiferroelectric for the infinite size system and thus has a double degenerate groundstate.
\label{fig:ring1D}}
\end{figure}
\begin{figure}
  \includegraphics[width=3.4in]{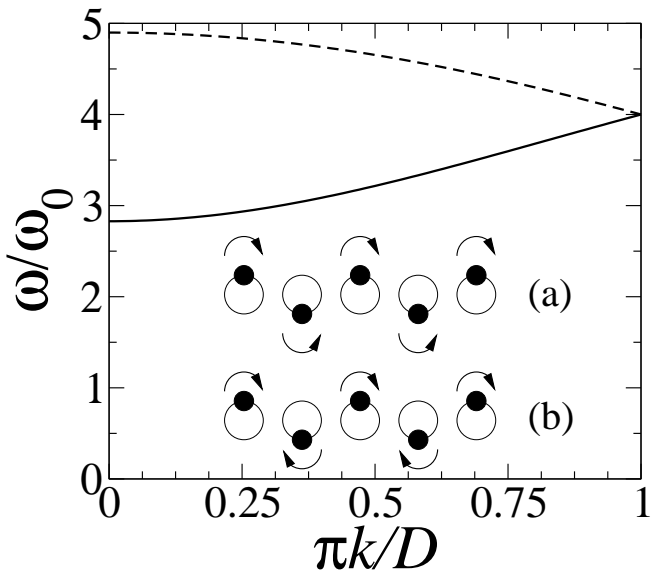}\\
  \caption{Normal modes of the 1D ring array with gaps of (solid line,
inset (a)) $2\sqrt{2}\omega_0$ and
  (dashed line, inset(b)) $2\sqrt{6}\omega_0$, where $\omega_0 =
\sqrt{e^2/m^* D^3}$}\label{fig:1Dmodes}
\end{figure}

\section{The model}

\label{sec:model} We consider a one dimensional array
of singly charged narrow quantum rings with radius $R$ and
center-to-center separation of $D$ (Fig.\ref{fig:ring1D}). The width
of each ring is much smaller than its inner radius so that 
we need only consider the
one dimensional movement of the electron around
the ring. While the rings are isolated from each other so there is no
charge transfer between rings, the rings still interact electrostatically
with their nearest neighbors.\cite{limits}
While in principle the Coulomb interaction is long range, we assume
that there is sufficient screening that next nearest neighbor interactions
can be neglected.

We also neglect tunneling between rings.   This is reasonable since
the physical realizations discussed above produce rings that are not
in tunneling contact.  Tunneling will be exponentially suppressed with
$D$, so that for spatially distinct rings it will be minimal.  

 All the phenomena explained in this article only will appear in 
experiments if the electrons do not lose their quantum mechanical
phase, i.e. the ring's perimeter has to be smaller than the electron's
coherence length ($ 2\pi R < L_\phi$ ) and the temperature has to
be lower than the dephasing temperature ($  T<T_\phi $).  

 In each ring the confinement energy of the
electron scales as $E_q=\hbar^2/2m^*R^2$; this energy opposes
localization of the wavefunction in the ring. The
inter-ring Coulomb repulsion, which scales as $E_c=e^2/D$, tries to
localize the wavefunction.  Electrons are repelled from 
regions of the ring where it is too close to the charges on 
neighboring rings.
The competition between these two physical scales
creates a \textit{quantum phase transition} in the array from
a localized state to extended state as we will see below.

\section{Classical results}\label{sec:classical} Before
solving a quantum mechanical problem it is often helpful to look
at the similar classical case which is usually easier to solve.
Below we consider two classical models, one in which the classical charges
are treated as ideal points, and the second in which they are treated as a 
continuous fluid.   

\subsection{Classical point charges}

 The classical model considers one charged point particle per ring
 with only
nearest neighbor Coulomb interaction. Unlike the quantum mechanical
case there is only one energy scale in the classical problem which
is the Coulomb energy $E_c=e^2/D$. The energy of a 1D array is given
by $U_{1D}=\sum_{i=1}^{N}e^2/|{\vec r}_i(\theta_i)-{\vec
r}_{i+1}(\theta_{i+1})|$, where $\theta_i$ is the location of the
$i$-th electron as measured from horizontal axis. In the dipole
approximation we can write this as :
\begin{eqnarray}\label{eq:1DU}
U_{1D}-U_0&\approx& {\epsilon^2 e^2\over 2D}
\sum_i\left[3\cos 2\theta_i+\cos(\theta_i-\theta_{i+1}) \right. \nonumber \\
&-& \left. 3\cos(\theta_i+\theta_{i+1})\right]  \nonumber \\
 &=& {\epsilon^2 e^2 \over D}\sum_i\left[{\vec s}_i\cdot {\vec
s}_{i+1}+{3\over 2}({\hat D}\cdot({\vec s}_i-{\vec s}_{i+1}))^2\right].
 \nonumber \\
\end{eqnarray}
where $\epsilon\equiv R/D$ and $U_0$ is a constant, $U_0\equiv {N
e^2\over D}(1+{\epsilon^2\over 2})$. In the second expression we
identify the position of each charge by a vector ${\vec s}_i$ in
the 2D plane pointing from the center of the $i$-th ring to the
charge on that ring. The unit vector ${\hat D}$ lies on the horizontal
axis.

 The $\cos 2\theta$ (or $({\hat D}\cdot {\vec s})^2$) term explicitly
 breaks the rotational
symmetry, driving the system from XY to Ising-like behavior. The
Heisenberg term in the last line of equation (\ref{eq:1DU}) drives
the system ferroelectric at zero temperature while the second
and larger term favors states where neighbors point in opposite
direction. Thus the system at zero temperature orders in an
antiferroelectric (AFE) pattern (Fig.\ref{fig:ring1D}) in one
dimension. Our numerical Monte Carlo simulations of the exact Coulomb
interaction also verifies the existence of such a minimum energy
configuration in the classical finite size arrays.
\begin{figure}
  \includegraphics[width=3.4in,height=2.7in]{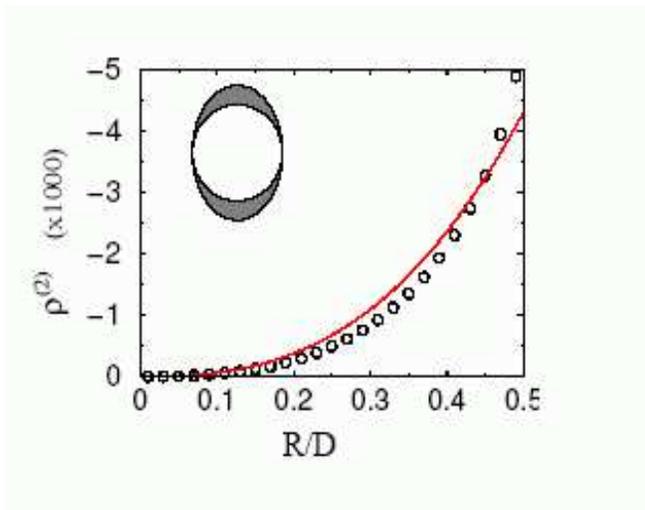}\\
  \caption{A plot of the second fourier amplitude of the classical
charge distribution on a ring in a 1D horizontal array. The circles
are numerical results, the solid line is a scaled plot of
eq.(\ref{eq:1Dfluid}). Scaling is required since the analytic result
neglects all higher fourier modes. Inset: a sketch of the charge
distribution that corresponds to this Fourier mode. Note that the
symmetry of the array is not broken by the charge
distribution.}\label{fig:fluidplot1D} \end{figure}


We can examine the stability of the AFE state by finding the higher
energy modes of the system.  We expand the energy function
(\ref{eq:1DU}) to quadratic order in displacement angle around the
AFE configuration using $\theta_i=(-1)^i{\pi\over 2}+\alpha_i$. The
AFE configuration has a basis with two sites so we find two independent
normal modes with frequencies : \begin{equation}\label{eq:1Dmode}
\omega^{(1D)}_{\pm}(k)=2\omega_0\sqrt{4\pm 2\cos{kD\over 2}} .
\end{equation} Where $\omega_0\equiv \sqrt{e^2/m^*D^3}$. Both the
modes are gapped since the Ising-like term provides the harmonic
restoring force at each site. The modes are shown in
Fig.(\ref{fig:1Dmodes}). Normal modes are found to be independent
of the ring radius.

\subsection{Classical charge fluid}
Another interesting classical limit of our ring problem is when
there is a classical \textit{self-interacting} fluid of charge on
each ring while the nearest neighbor fluids are still interacting
with each other. 

 To find the minmum energy distribution of charge density on each ring we define
an angular dependent charge density $\rho_i(\theta_i)$ on each ring where:$\int\rho_i(\theta_i)d\theta_i=1$.  We
are looking for the minimal solution to the variational quantity :
\begin{equation}\label{eq:fluid}
I={1\over 2}\int d\theta \int d\theta' \sum_{\langle ij\rangle}{\rho_i(\theta)\rho_j(\theta')\over |{\vec
r}_i-{\vec r}_j|}+\lambda\sum_i\int d\theta\rho_i(\theta).
\end{equation}
For a 1D ring this expression is divergent due to self energy. We can regularize this in several ways. One
method is to introduce a short distance cutoff $\zeta$ to the Coulomb interaction, discretize the integral
equation and then solve the problem numerically. 
 An approximate analytic solution can then be obtained
by Fourier expanding the distribution, keeping only the first three modes.

 For 1D array with periodic
boundary condition we find that the amplitude of the 
non-trivial Fourier mode as a function of $\zeta$ and
$\epsilon\equiv R/D$ for $\rho$ is given by :
\begin{equation}\label{eq:1Dfluid}
\hat{\rho}^{(2)}\approx {-3\pi\epsilon^3(2-5\zeta^2)\over 4(-2+4\log(\epsilon/\zeta))} .
\end{equation}
We compare this analytic result with the numerical diagonalization of (\ref{eq:fluid}) in
Fig.(\ref{fig:fluidplot1D}). As we can see in Fig.(\ref{fig:fluidplot1D}) the minimum energy configuration of the
1D array of charge fluid does not break the up-down symmetry of the system.

\section{Quantum results for $B=0$.} 
\label{sec:quantum} 

At first glance the
quantum mechanical wavefunction of a charged particle resembles the
classical charge fluid.  Although the wavefunction does not have
a self-interaction, the quantum particle has a kinetic energy which
opposes localization, making the analogy to charge fluid even
stronger.  Unlike the classical case, the quantum problem has
 two competing energy scales:
the quantum
kinetic energy, $E_q$, preventing localization  and the Coulomb
interaction energy, $E_c$, trying to force the charges away from
each other. At $E_q\ll E_c$ we expect charge localization on each
ring and at $E_q\gg E_c$ we expect no localization of charge. However
it is not \textit{a priori} obvious whether the charge localization
(system polarization) breaks symmetry or not, whether this
localization is a smooth function of the external parameters, and
if it is a phase transition, what is the exact nature of this transition.

\subsection{Variational calculation}

As a first step we can use a simple variational wavefunction to
find the polarization behavior of the system in ground state.
The
dimensionless hamiltonian of an array of one dimensional rings with
radius $R$ is given by:
\begin{equation}\label{eq:hamiltonian}
{\hat H}= -\sum_i {\partial^2\over\partial\theta_i^2}+\delta\sum_{\langle ij\rangle}{1\over |{\vec
r}_i(\theta_i)-{\vec r}_j(\theta_j)|},
\end{equation}
where $\delta=E_c/E_q$ is the interaction strength and the energy is measured in units of
$E_q=\hbar^2/2mR^2$.

To find the ground state energy of the 
 1D array we employ a simple
ansatz for the wavefunction of each sublattice : $\psi_A(\theta)={\sqrt{1-y^2}\over\sqrt
2\pi}+{y\over\sqrt{2\pi}}\cos(\theta-\phi)$ and $\psi_B(\theta)={\sqrt{1-y^2}\over\sqrt
2\pi}-{y\over\sqrt{2\pi}}\cos(\theta-\phi)$ alternately.
 The ground state values of $y$ and $\phi$ is obtained
by minimizing the energy (\ref{eq:hamiltonian}) using dipole approximation for Coulomb interaction we find:
\begin{equation}\label{eq:1Dalalytic}
y(\delta,\epsilon)=\cases{
{1\over 4}\sqrt{11-{4\over \delta\epsilon^2}} & for $\delta\geq\delta_c(\epsilon)$\cr
0&  $\delta<\delta_c(\epsilon) $ }
\end{equation}
and $\phi= \pi/2$, where  the critical value of the interaction is given by
$\delta_c(\epsilon)={4\over 11}\epsilon^{-2}$. 
To find out the degree of polarization we define the \textit{staggered
polarization} vector as:
\begin{equation}\label{eq:1Dstp}
{\vec P}_s=\sum_i (-1)^i\int d\theta |\psi_i(\theta)|^2 {\vec r}_i(\theta) .
\end{equation}
Using variational results the staggered polarization of the system is as follows:
\begin{equation}\label{eq:1DstpResults}
{\vec P}_s(\delta,\epsilon)=\cases{
{1\over 8}\left({4+5\, \delta\epsilon^2\over 2\delta\epsilon^2}\right)^{1\over 2}
(11-{4\over \delta\epsilon^2})^{1\over 2}{\hat D}_\bot & $ \delta\geq\delta_c $ \cr
0 & $ \delta\leq\delta_c$} .
\end{equation}
Where ${\hat D}_\bot$ is the unit vector perpendicular to the common axis of the rings (Fig.\ref{fig:ring1D})
and $\delta_c$ is defined in equation (\ref{eq:1Dalalytic}). 

 As we can see variational calculation
suggests that the ground state of the 1D array of rings
antiferroelecrically polarizes in perpendicular direction at high
interaction strengthes while at lower values the wavefunctions are
not localized, hence the system has no polarization.
The validity
of this result will be confirmed in next sections using more exact
and reliable methods of calculation.
\begin{figure}
  \centering
  \includegraphics[width=3.5in]{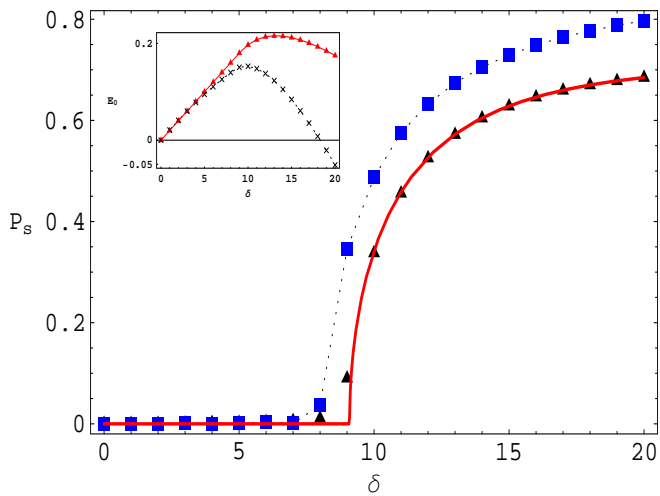}\\
  \caption{A comparison of numerical and analytical calcula-
tions of the staggered polarization and energy as a function of $\delta$ in a 1D quantum ring array obtained in
the Hartree approximation. The numerical results are for the case include Fourier modes $|m|\leq 1$ (triangles)
and $|m|\leq 6$ (boxes). The solid line is the analytic result assum- ing $|m|\leq 1$. The quantity $\delta$ is
a measure of the competition between the Coulomb interaction and the quantum kinetic
energy \label{fig:hartree1D}}
\end{figure}

\subsection{Hartree Approximation} The rings 
considered here are well
separated with exactly one electron
on each ring.  Under this condition and because of strong Coulomb
repulsion the effect of inter-ring transfer of electrons and overlap
of wavefunctions is small. We
can therefore neglect the inter-ring transfer from our calculations. Since
without overlap the electrons do not have any exchange interaction,
the Hartree approximation is  exact for this problem.\cite{Hartree}

 We can decompose the wavefunction in each ring into a limited
 number of Fourier modes,
$\psi_i(\theta)=\sum_{n=-n_0}^{n_0}c_ne^{in\theta}$, and then solve
the system numerically in the Hartree approximation. We impose the
periodic boundary conditions on the array and by an iterative
self-consistent method we find the ground state wavefunction of the
rings. In Fig.(\ref{fig:hartree1D}) we can see the numerical results
of the polarization and energy change of the 1D array of rings for
different number of Fourier modes using exact Coulomb interaction
and also its agreement with the variational calculation when we
restrict the number of Fourier modes to $n\in \{-1,0,1\}$. The
results are little changed when we increase the number of Fourier
modes mostly in the high coupling regime.

 All the above results suggest that there is transition
from unpolarized to polarized state at zero temperature by changing
the coupling.  By looking at the behavior of polarization when the
number of Fourier modes increases we realize that this transition
tends to be sharper and sharper for higher number of Fourier modes
suggesting a true phase transition in the system. If true, this
transition would be a sudden change of ground state of the quantum
system at zero temperature, known as 
\textit{quantum phase transition}.\cite{sachdev}
We demonstrate that this is the case and determine the 
universality class of the transition in the next section using the
Monte Carlo simulation.

\subsection{Monte Carlo simulation} It is well-known that we can
write the quantum partition function of a quantum system,
$Z=Tre^{-\beta{\hat H}}$ as the sum over all paths taken by the
system in imaginary time defined by the scale $\hbar\beta$. If
the quantum system is D-dimensional then the partition function
will look like the path integral of a D+1-dimensional classical
system in which the extra dimension is the time direction
$0<\tau<\beta\hbar$. At zero temperature $\beta\rightarrow\infty$
the classical system is truly D+1-dimensional.
One can derive an
effective Hamiltonian for such a classical system from the quantum
Hamiltonian using a complete set of basis states.  In this classical
system the parameters of the quantum system (in our case $\delta$)
is a control knob like temperature. We can use Monte Carlo simulation
of such a classical system and find out the universal behavior of
the quantum system.

 To develop a 1+1-dimensional classical theory for our 1D ring array
 we first stagger
the order parameter, $\theta_i\rightarrow (-1)^i\theta_i$ so that
we can analyze the Monte Carlo results easily. We also use dipole
approximation for the Coulomb interaction. Consequently we can write
the Hamiltonian of the system as:
\begin{equation}\label{eq:1DQH}
\hat{\cal H}={E_c\over 2}\sum_{j=1}^N(-i{\partial\over\partial\theta_j})^2-E_J\sum_{j=1}^N \hat{V}.
\end{equation}
In which $E_c=\hbar^2/ mR^2$ and $E_J=e^2\epsilon^2/2D$. 
The standard derivation\cite{AnalyticContinuation} 
using the Villain approximation\cite{Villain,JoseKad}
tells us that the 1+1D classical partition function equivalent to the 1D interacting quantum ring array at zero
temperature is:
\begin{eqnarray}\label{eq:1+1D}
Z&\propto& \int {\cal D}\theta(\tau) \prod_{a=1}^{\cal N}\exp{\Big \{}{\hbar\over
E_c\delta\tau}\sum_{k=1}^N\cos[\theta_k(\tau_{a+1})-\theta_k(\tau_a)]
\nonumber \\
 &+&{\delta\tau E_J\over\hbar}\sum_{k=1}^N V_k(\tau_a){\Big\}} \ .
\end{eqnarray}
Where ${\cal D}(\theta)\equiv \prod_{a=1}^{{\cal N}}D\theta(\tau_a)$ and
\begin{eqnarray}\label{eq:pot}
V_k(\tau_a)&=& 3\cos[\theta_{k}(\tau_{a})-
\theta_{k+1}(\tau_a)]+\cos[\theta_{k}(\tau_{a})+\theta_{k+1}(\tau_a)] \nonumber\\
&-&3\cos[\theta_{k}(\tau_{a})-\theta_{k+1}(\tau_a)]
\cos[\theta_{k}(\tau_{a})+\theta_{k+1}(\tau_a)]. \nonumber \\
\end{eqnarray}
The parameter $\tau$ has the dimension of time and ${\cal N}\delta\tau=\beta\hbar$. It can be shown that the
field $\theta(x,\tau)$ obeys the periodic boundary condition , $\theta(\tau+\beta\hbar)=\theta(\tau)$ 
(\cite{pbc}).
We will also assume periodic boundary condition in space direction all over the simulation.

 By defining
the spin vector ${\vec S}_i=(\cos\theta_i,\sin\theta_i)$ we can
interpret Eq.(\ref{eq:1+1D}) as a two dimensional classical spin
model. Our early calculations suggested that the system of 1D rings
has a transition from the unpolarized to the  AFE state. In this
classical analogue because we have already staggered the order
parameter we expect to see a transition from unpolarized to
ferromagnetically polarized state (FE). Close to this transition the
spatial variation of the order parameter ${\vec S}$ is smooth so
we can approximate (\ref{eq:pot}) as follows:
\begin{equation}\label{eq:appot}
V_k(\tau_a)\approx 3\cos(\theta_k(\tau_a)-\theta_{k+1}(\tau_a))-2\cos 2\theta_k(\tau_a).
\end{equation}
Using the above potential finally the classical partition function looks like:
\begin{equation}\label{eq:apph}
Z\propto\int {\cal D}\theta(\tau)\exp{\Big\{}K\sum_{\langle ij\rangle}\cos(\theta_i-\theta_j)-{2K\over
3}\sum_i\cos 2\theta_i{\Big\}}.
\end{equation}
Where i and j run over an infinite 2D square lattice and we have determined $\delta\tau$ to identify the two
couplings in eq.(\ref{eq:1+1D}) as $K=\sqrt{3E_J\over E_c}$ . Equation (\ref{eq:apph}) is a 2D XY model with a
symmetry breaking field which is $2/ 3$ the XY coupling. Our Monte Carlo analysis shows that this model
has a continuous phase transition. The order parameter of this system is the total magnetization
density equivalent to the total staggered polarization of the 1D ring array:
\begin{equation}\label{eq:m_Ps}
{\vec m}\equiv\langle {\vec S}\rangle\longleftrightarrow {\vec P}_s.
\end{equation}
Where the average on the left hand side is the thermodynamic average over the infinite size lattice.

\begin{figure}
\centering
\includegraphics[width=3.4in]{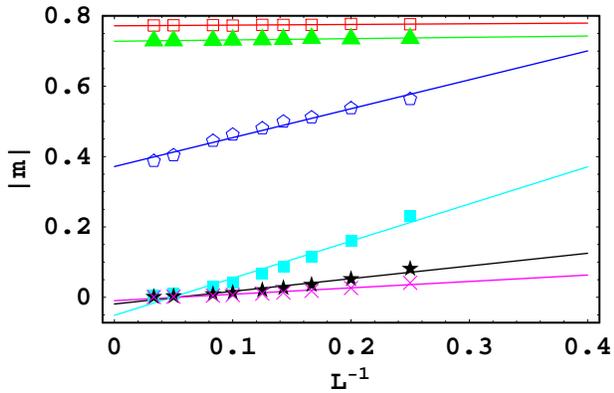}\\
\caption{Extrapolation of the total magnetization density of the 1+1D classical system to infinite size at
couplings $K/K_c(\infty)$=1.45(empty boxes), 1.32(triangles), 1.03(polygons), 0.74(filled boxes), 0.45(stars)
and 0.16(crosses). The solid lines are linear fit to each set of data.}\label{fig:mextrapolate}
\end{figure}

 The
fluctuations in this system is controlled by $K$ which is the analogue of $1/T$ in real classical systems. We can
measure the analogue of specific heat of the system using:
\begin{equation}\label{eq:cv}
\tilde C_v={1\over N^2}(\langle E^2\rangle-\langle E\rangle^2)
 \end{equation}
\begin{figure}
\begin{center}
  \includegraphics[width=3.4in]{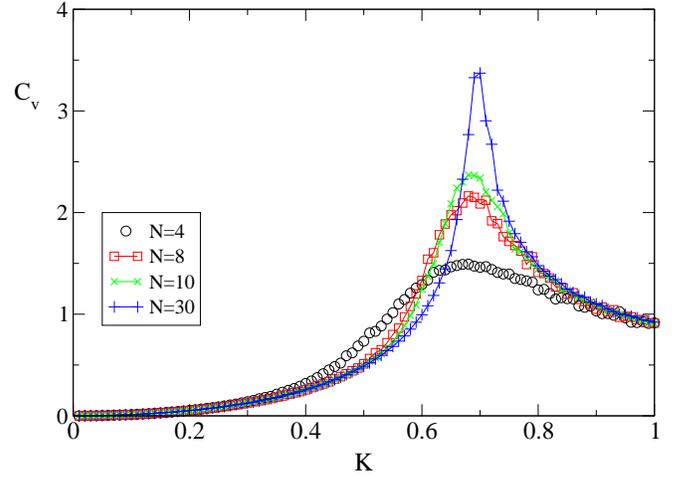}\\
  \caption{Monte Carlo results of the $\tilde C_v=\overline{\Delta E^2}/N^2$ for different system sizes.
  The system is a 1+1D classical equivalent of 1D quantum ring array at zero temperature.}\label{fig:cv}
  \end{center}
\end{figure}
\begin{figure}
\includegraphics[width=3.4in]{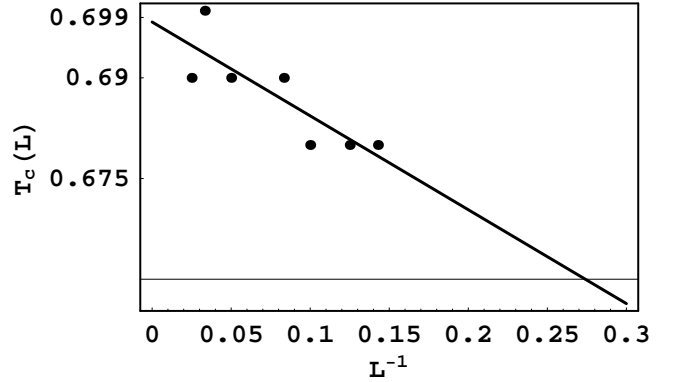}\\
\caption{Plot of critical coupling $K_c(L)$ at different system sizes taken from the $C_v$ plots. The solid line
is a linear fit to the data indicating $K_c(\infty)\approx 0.699$.}\label{fig:maxcvplot}
\end{figure}
in which $\langle . \rangle$ is the average over an ensemble and
$E$ is the total energy
 of the $N\times N$ system. This quantity diverges at the critical point of the
 infinite system undergoing a continuous phase transition.
Fig.(\ref{fig:cv}) shows the change of the specific heat of our
1+1D system in terms of the parameter $K$ for different lattice
sizes. As we can see at $K_c$ the peak gets sharper and sharper
with increasing lattice size $L$.  An extrapolation of the point
of the maximum of $C_v$, $K_c(L)$ to $L^{-1}=0$ determines the
approximate critical point of the infinite lattice(Fig.\ref{fig:maxcvplot})
. Also an extrapolation of $m(L,K)$ for different values of $K$ in
Fig.(\ref{fig:mextrapolate}) shows that a real continues phase
transition happens in the infinite size system.

 The effective classical system derived here does not fully explain
 all the physical aspects of
the 1D quantum system mainly because of the approximations used to
derive the path integral. However, we believe that, close to the critical
region, these approximations do not play any role in the general
behavior of the system and the universality class remains unchanged.
Hence using the finite size scaling method we can determine the
critical exponents of the classical system and determine the
universality class of the actual quantum system.

\subsubsection{Finite size scaling of the 1+1D system}
\begin{figure}
  \centering
  \includegraphics[width=3.4in]{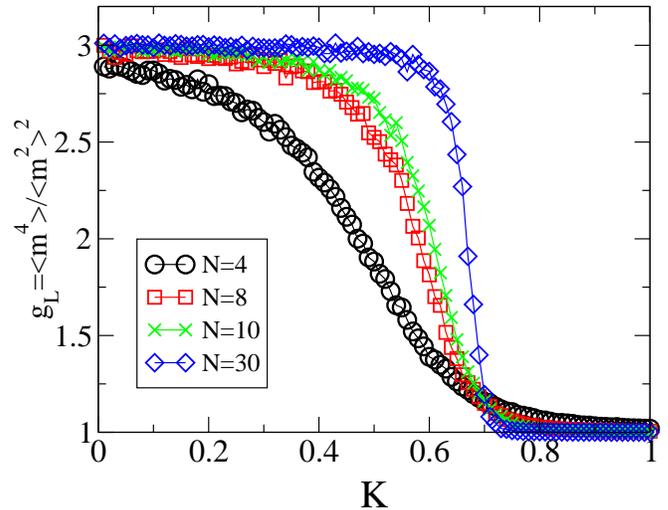}\\
  \caption{Plots of Binder ratio for different system sizes.
  The behavior is sharper at larger sizes.}\label{fig:binder}
\end{figure}

\begin{figure}
\includegraphics[width=3.4in]{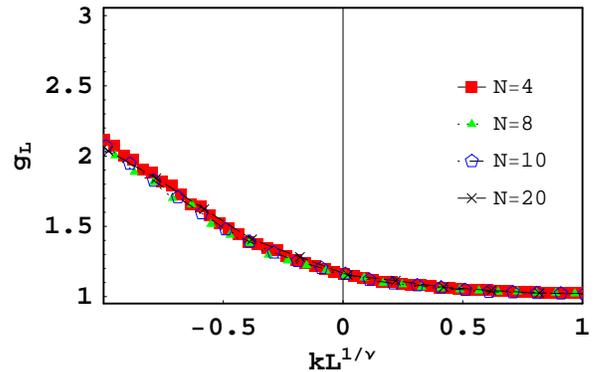}\\
\caption{Collapse of Binder plots at the critical region.
The best collapse is obtained for
$\nu=1.01\pm .01$. \label{fig:bindercollapse}}
\end{figure}

\begin{figure}
\includegraphics[width=3.4in]{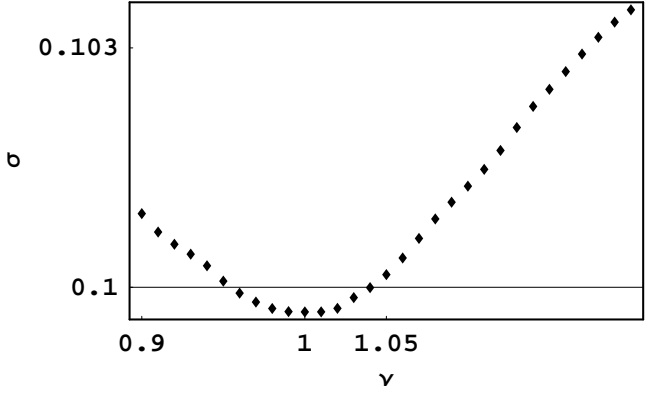}\\
\caption{Standard deviation of the set of 
scaled plots of Binder ratio for different exponents. The case for
$\nu=1$ is the best choice which is plotted in Fig.(\ref{fig:bindercollapse}).}\label{fig:nuplot}
\end{figure}

One of the best parameters for examining the phase transition and find the universal exponents with finite size
scaling is the dimensionless Binder ratio:
\begin{equation}\label{eq:BinderDef}
g_L={\langle m^4\rangle\over \langle m^2\rangle^2}
\end{equation}
defined for a system with size $L$. In the disordered phase $K<K_c$ the correlation length $\xi$ is finite so
for $L\gg\xi$ the distribution of $m$ is Gaussian around $m=0$ with the width $\sim N^{-1/2}\sim L^{-d/2}$ so
$g\rightarrow 0$. On the other hand for $K>K_c$ where $\langle m\rangle$ is finite, $g_L$ approaches a constant
as $L\rightarrow\infty$.
The variation of $g_L$ with $K$ becomes sharper and sharper as $L$ increases, however
all the $g$'s cross at the transition point $K_c$.
The variation is given by the following finite size scaling
function:
\begin{equation}\label{eq:BinderSc}
g_L(K)=\widetilde{g}(L^{1/\nu}(K-K_c)).
\end{equation}
Where $\widetilde{g}$ is a scaling function which depends on $L$ and $K$ only in that particular form. By using
the finite size data we can try to find a data collapse and by calculating the 
standard deviation find the best
exponent $\nu$ fitting to the collapsed function. Fig(\ref{fig:binder}) shows the Binder ratio for different
lattice sizes. Fig.(\ref{fig:bindercollapse}) shows the collapsed data and Fig.(\ref{fig:nuplot}) shows the best
exponent is $\nu=0.99\pm 0.01$.  The error is estimated from the mesh of the numerical calculation.

\begin{figure}
\includegraphics[width=3.4in]{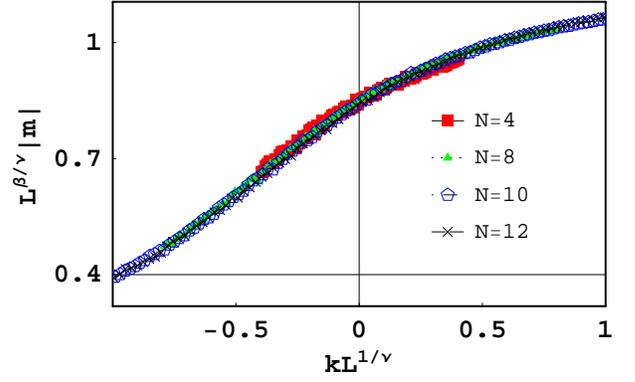}\\
\caption{Collapse of scaled magnetization data sets for $\nu=1$ and $\beta=1/8$ in the critical
region.\label{fig:betacollapse} }
\end{figure}

\begin{figure}
\includegraphics[width=3.4in]{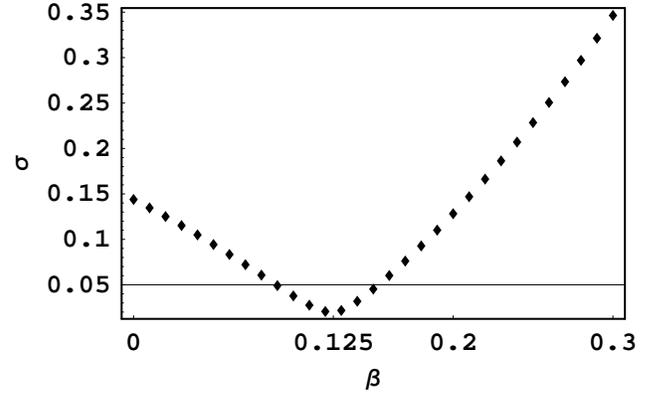}\\
\caption{Standard deviation of the set of scaled plots of magnetization for fixed $\nu=1$ and different exponent
$\beta$. The case for $\beta=1/8$ is the best choice which is plotted in
Fig.(\ref{fig:betacollapse}).}\label{fig:betaplot}
\end{figure}

 The scaling for the order parameter $|m|$ is:
\begin{equation}\label{eq:orderSc}
m=L^{-\beta/\nu}X^0(L^{1/\nu}(K-K_c)).
\end{equation}
Where $X^0$ is a function of $x=L^{1/\nu}(K-K_c)$ only and $\beta$ is one of the universal scaling exponents of
the system. To determine the universal exponent $\beta$ we plot $L^{\beta/\nu}$ vs. $x$ for different
sizes. 
Fig.(\ref{fig:betacollapse}) and (\ref{fig:betaplot}) show the the collapse of different data sets and the
standard deviation for different 
exponents respectively which shows the
best estimation is $\beta=0.125\pm .005$, or $1/8$.
\begin{figure}
\centering
\includegraphics[width=3.4in]{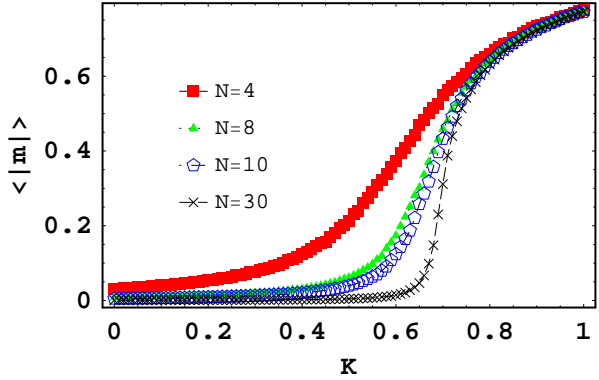}\\
\caption{Monte Carlo results of average magnetization of the 1+1D classical system for different system sizes.
The change of magnetization tends to be sharper as the system size grows.}\label{fig:abs}
\end{figure}
\begin{figure}
\centering
\includegraphics[width=3.4in]{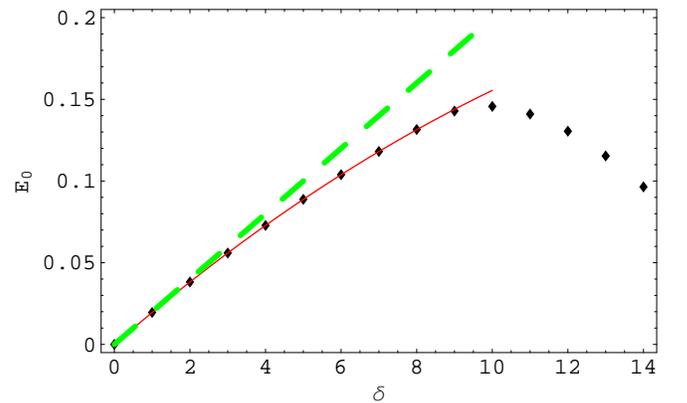}\\
\caption{Plot of the energy of interacting quantum ring array when all the wavefunctions are constant around the
ring (dashed) or all are in the form of $\psi_d=a+b\cos2\theta$(solid). The points are the actual results coming
out of the numerical Hartree calculation indicating that $\psi_d$ is the selected behavior for
$\delta<\delta_c\approx 8$.}\label{fig:anlenergy}
\end{figure}

\subsubsection{Universality Class} The universal exponents extracted
from the finite size data indicate that our 1+1D classical XY model
in the symmetry breaking field is in the universality class of 2D
classical Ising model, hence the nature of quantum phase transition
of our 1D quantum ring array is Ising-like. The Coulomb repulsion forces
the electrons to alternate staying on the top and bottom of the rings.
However the quantum kinetic energy tries to avoid localization.
This kinetic energy causes the electrons to tunnel from top into
bottom of the ring hence destroying the antiferroelectric order.
This ordering behavior shows up in the probability  distribution
on each ring. Fig.(\ref{fig:anlenergy}) shows the energy of each
electron with the wavefunction $\psi_d(\theta)=x+y\cos(2\theta)$
compared to when the wavefunction is a constant all around the
ring. The wavefunction $\psi_d$ has two maxima on top and bottom of
the ring which means the electron is fluctuating up and down. As
we can see by increasing the coupling the lower energy state selected
by the exact Hartree calculations (dots) gradually matches $\psi_d$
instead of the constant wavefunction. This behavior persists in a range
of couplings close but smaller than the critical coupling i.e. in
disordered region $\delta<\delta_c$. Needless to say that after
transition point the ground state wavefunction is no longer $\psi_d$
and the system starts to excite more angular momentum eigenstates
(Fourier modes).

All the above discussion suggests that the nature of the antiferroelectric transition is not just the simple 2D
Ising but is similar to \textit{1D transverse field Ising}(TFI) which has a quantum phase transition at zero
temperature in the same universality class as 2D Ising. We can develop an effective 1D TFI hamiltonian for our
ring array in the dipole approximation. In this approximation we can write down the hamiltonian(\ref{eq:1DQH})
as follows :
\begin{eqnarray}\label{eq:1DTFI}
{\hat H}&=&{\hat H}_0+{\hat V} \nonumber\\
 {\hat H}_0&=&\sum_{i}\left[(-i{\partial\over\partial\theta_i})^2
 +3\delta\epsilon^2\cos2\theta_i \right] \nonumber\\
 {\hat V}&=&-\delta\epsilon^2\sum_{i}\left[3\cos(\theta_i+\theta_{i+1})+\cos(\theta_i-\theta_{i+1})\right].
\nonumber \\
\end{eqnarray}
\begin{figure}
\includegraphics[width=3.4in]{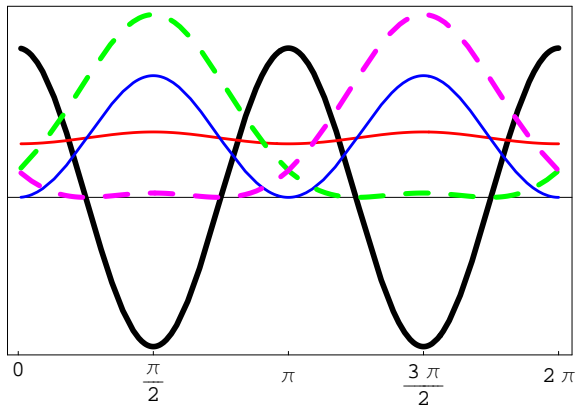}\\
\caption{Plot of the $\cos2\theta$ potential around a ring (the thick solid line), the ground and first excited
state of this potential (thin solid lines) coming out of a simple numerical Schrodinger equation solver and the
up and down states (dashed lines) constructed from the two eigenstates(see appendix). The scale of the potential
is exaggerated for easier comparison.}\label{fig:pot}
\end{figure}
The $\cos2\theta$ term in hamiltonian ${\hat H}_0$ has two minima
at top and bottom of the ring. Fig.(\ref{fig:pot}) shows the potential
and the two lowest energy states of it with energies $E_0<E_1$. The
rate of tunnelling from top to bottom or vice versa is determined
by $\Delta\equiv E_1-E_0$. The potential ${\hat V}$ tries to align
the electrons hence it acts like the Ising interaction. A more
rigorous derivation using the Holstein-Primakov bosons\cite{bosons} shows that
the projection of the hamiltonian ${\cal H}$ into the subspace of
the ground and first excited states of ${\hat H}_0$ can be written
as (see appendix):
\begin{equation}\label{eq:TFIhamiltonian}
{\hat H}\approx \Delta\sum_{i=1}^{N}\sigma^x_i-J\sum_{i=1}^{N}\sigma^z_i\sigma^z_{i+1}.
\end{equation}
In which $\sigma$'s are Pauli spin matrices and $J=8\delta\epsilon^2$. Numerical diagonalization of ${\hat H}_0$
tells us that $\Delta\approx 1-0.1\delta\epsilon^2$ for small $\epsilon$. 

 Close to transition the
Coulomb repulsion is not strong enough to excite the electrons to
higher states, consequently the TFI model in Eq.(\ref{eq:TFIhamiltonian})
is valid and indicated the nature of transition of the 1D ring
array.

\section{Quantum rings transitions for $B\neq0$}

\label{sec:Bfield}
Y. Aharanov and D. Bohm (AB) have predicted that the wavefunction
of an electron moving in a vector potential ${\vec A}(x)$ along the
path $C$ acquires a phase shift: \begin{equation}\label{eq:ABphase}
\Delta\Lambda={e\over\hbar c}\int_{C}{\vec A}\cdot{\vec dr}.
\end{equation} AB predicted that this phase shift can be observable. When
an electron is confined on a closed path like the case of charged
ring threaded by magnetic flux $\phi$ the phase shift after one
$2\pi$ rotation would be :

\begin{equation}\label{eq:ABflux}
\Delta\Lambda={e\over\hbar c}\oint {\vec A}\cdot{\vec dr}=\phi/\phi_0.
\end{equation}
Where $\phi_0=hc/e\simeq4.135\times10^{-7} G.cm^2$ is the quantum of flux. The phase shift above has been
observed in numerous experiments and different devices 
including the experiments of persistent current and
excitons in quantum rings.  
In this section we show 
how magnetic field changes the behavior of polarization.

The Hamiltonian of an electron in 1D
a ring threaded by a constant uniform magnetic field $B{\hat z}$ (the ring
is in the x-y plane) is:
\begin{equation}\label{eq:gaugehampolar}
{\hat H}={\hbar^2\over 2mR^2}(i{\partial\over\partial\theta}+{\phi\over\phi_0})
\end{equation}
in which the choice of gauge: ${\vec A}={B\over 2}(-y,x,0)$, 
the momentum is in polar coordinates:${\hat p}=-i{\hbar\over R}{\partial\over\partial\theta}$ and
the eigenfunctions are periodic: $\psi(\theta+2\pi)=\psi(\theta)$. The eigenenergies of (\ref{eq:gaugehampolar})
will be:
\begin{equation}
E_n={\hbar^2\over 2mR^2}(n-\phi/\phi_0)^2
\end{equation}
 in which $n$ is an integer. By changing the gauge ${\vec
A}\rightarrow {\vec A}-\nabla\Lambda$ wavefunctions undergo a phase change $\psi\rightarrow e^{{ie\over\hbar
c}\Lambda}\psi$. For example we can use the gauge transformation with the choice of $\Lambda=(BR^2/2)\theta$ to
remove the vector potential from (\ref{eq:gaugehampolar}) but at the same time we have to shift the phase
of the wavefunctions to $e^{-i{\phi\over \phi_0}\theta}\psi$. As the result the eigenfunctions change:
\begin{equation}\label{eq:gaugefxn}
\psi_n(\theta)={1\over\sqrt{2\pi}}e^{i(n-\phi/\phi_0)\theta}.
\end{equation}
This eigenfunction however has a 
different boundary condition than the previous one :
\begin{equation}\label{eq:gaugeE}
\psi(\theta+2\pi)=e^{2\pi i\phi/\phi_0}\psi(\theta)
\end{equation}
\begin{figure}
\centering
\includegraphics[width=3.4in]{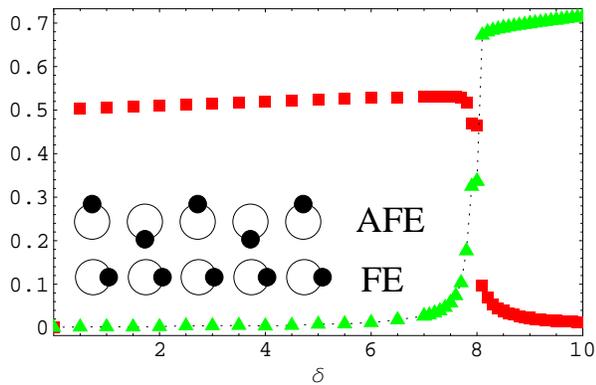}
 \caption{Results of numerical Hartree calculations of the polarization of a 1D quantum ring array
threaded by half-flux quantum $\phi/\phi_0=1/2$. For $\delta<\delta_c$ the system displays a longitudinal
ferroelectric (FE) polarization (squares), while for $\delta>\delta_c$ an  antiferroelectric (AFE) polarization
(triangles) is observed.  
}\label{fig:polB1D}
\end{figure}
 but as one physically expects the eigenenergies are not changed. The ground state energy of the interacting
quantum ring array in the magnetic flux $\phi$ can be written as:
\begin{eqnarray}\label{eq:energyB}
E_0(\delta,\phi)&=&\sum_{n=-n_0}^{n_0}|c_n|^2(n-\phi/\phi_0)^2+\nonumber \\
&+&\delta\sum_{\langle ij\rangle}\int d\theta d\theta'{|\Psi_0(\theta)|^2 |\Psi_0(\theta')|^2\over |{\vec
r}_i(\theta)-{\vec r}_j(\theta')}
\end{eqnarray}
\begin{figure}
\centering
\includegraphics[width=3.4in]{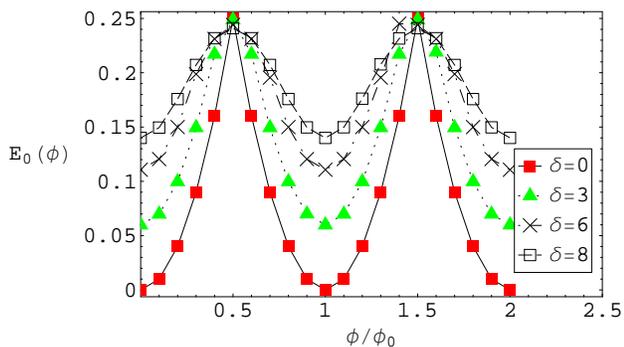}
\caption{Plots of ground state energy of the interacting quantum ring array in the external magnetic flux
threading each ring for different couplings $\delta$. The physics is periodic because of Aharanov-Bohm induced
phase that is proportional to the flux.}\label{fig:enBplot}
\end{figure}
where $\Psi_0(\theta)=e^{i\phi/\phi_0\theta}\sum_{n=-n_0}^{n_0}c_n
e^{in\theta}$ is the ground state
 wavefunction expanded in the free hamiltonian basis states. As we
 can see in
Eq.(\ref{eq:energyB}) the only part that is affected by the AB phase
is the kinetic energy and the potential energy is not sensitive to
the phase.  The energy calculated in 
Eq.(\ref{eq:energyB}) is periodic in $\phi_0$ since when
$\phi=\phi_0=1$ we
can rearrange the infinite sum and show that the value of the kinetic energy
is equal to its value at $\phi=0$. This is due to the well-known fact
that the physics of quantum rings 
does not change at integer flux quanta. For infinite $n_0$
this argument is true at any range of magnetic flux; in our
numerical calculations where we have used a finite number of Fourier
modes $E_0$ is periodic only in a finite range approximately given 
by  $0<\phi<n_0\phi_0$ in
Fig.(\ref{fig:enBplot}) we can see the periodic behavior of the
ground state energy of the ring array as magnetic flux changes.

\begin{figure}
\centering
\includegraphics[width=3.4in]{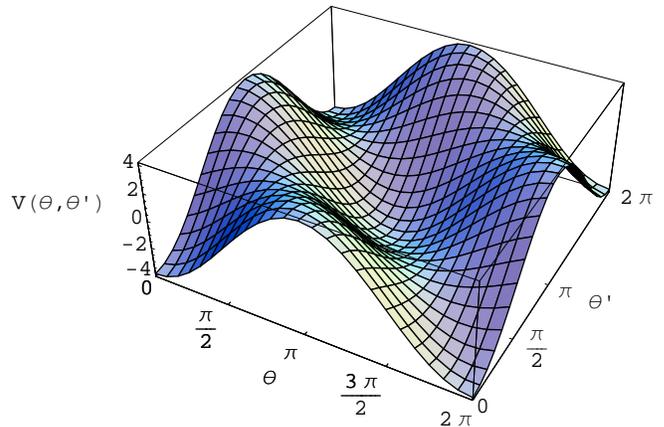}
\caption{Plot of the two-body potential function in Eq.\ref{eq:dipole+gauge}. The interacting rings in magnetic
field select the minimum of this potential for their ground state wavefunction at low couplings. The potential
is in units of $\delta\epsilon^2$.}\label{fig:bpotential}
\end{figure}

 The results of our numerical Hartree calculations
indicate that in a 1D ring array in which each ring is threaded by
a magnetic flux $\phi$ the polarization pattern changes from
unpolarized to ferroelectric at half-integer flux quantum.
Fig.(\ref{fig:polB1D}) shows the behavior of $P_x$ , the component
of the total polarization vector ${\vec P}=\sum_i\int_0^{2\pi}d\theta
|\psi_i(\theta)|^2{\vec r}_i(\theta)$ in the direction of the ring's
common axis at half flux quantum. This plot shows that at
$\delta<\delta_c$ the wavefunction has an unbalanced distribution
around each ring. However the total polarization vanishes at higher
values of interaction where the wavefunction distribution becomes
antiferroelectrially polarized in the array. The finite polarization
at small interaction strengths has a ferroelectric pattern which
is degenerate left or right. From this result we can see that the
physics of quantum ring arrays changes at half integer flux quantum.
The total staggered polarization in the ${\hat y}$ direction
perpendicular to the common axis of the rings starts to build up
at $\delta>\delta_c$ as in the case of no magnetic field.  We can
explain this phenomena of \textit{finite transverse polarization
due to magnetic field} in different approaches.

 We can use a simple perturbative discussion to understand this behavior
qualitatively. The Eq.(\ref{eq:1DTFI}) which is the dipole approximation of the total hamiltonian will modify in
presence of a magnetic flux as follows:
\begin{eqnarray}\label{eq:dipole+gauge}
{\hat H}&=&{\hat H}_0+{\hat V} \nonumber\\
 {\hat H}_0&=&\sum_{i}\left[(-i{\partial\over\partial\theta_i}-{\phi\over\phi_0})^2 \right.
 \left. +3\delta\epsilon^2\cos2\theta_i \right] \nonumber\\
 {\hat V}&=&-\delta\epsilon^2\sum_{i}\left[3\cos(\theta_i+\theta_{i+1})+\cos(\theta_i-\theta_{i+1})\right].
\nonumber \\
\end{eqnarray}
 In the above equation the kinetic energy hamiltonian has a degenerate ground state. For
example at half filling, $\phi/\phi_0={1\over 2}$, $n=0$ and $n=1$ levels are degenerate unlike the case of zero
magnetic field in which the ground state is unique and at $n=0$. By adding the symmetry breaking term
$\cos2\theta$ in the zero flux the electron gains enough energy to excite to the next higher level. This
excitation causes the electron to destroy any localization in the range where ${\hat V}$ is not strong enough
yet. However when there is a finite magnetic field the $\cos 2\theta$ can not lift the degeneracy between $n=0$
and $n=1$ levels in the range of small $\delta$'s. That is why in this case the ground state of ${\hat H}_0$
remains degenerate(Fig.\ref{fig:enBplot}). As long as $\delta$ is small the perturbative two body potential in
(\ref{eq:dipole+gauge}) can not excite the electron to higher levels and the kinetic energy of the electron
freezes. When this happens the electrons behave classicaly and choose a wavefunction that minimizes the
potential ${\hat V}$. In Fig.(\ref{fig:bpotential}) we can see a 3D plot of the two body potential $V$ in which
it has two stable minima at $(\pi,\pi)$ and $(0,0)$ indicating the preferred state of the quantum ring array at
low $\delta$ being the ferreoelectric right or left state.

It is surprising that the system orders ferroelectricallly along the chains.  This occurs because the
charge distribution is quite broad.   Using only the two lowest Fourier modes the charge distribution can 
be written in general as $\rho(\theta)= \left(1+\cos{(\theta-\theta_0)}\right)/2\pi $ , which nearly wraps around the ring.
If we change this to an (unphysical) flat charge distribution with an artificially varying width, the
FE state has a lower energy than the AFE state when the width $\Delta \theta \sim  .45 \pi$, while
for a triangular 
distribution the FE distribution is favored when $\Delta \theta \sim 0.6 \pi$.  
  While the exact transition
depends of course upon $\epsilon$, these calculations help explain why the FE phase wins out at half filling.

\section{Conclusions}
\label{sec:Conclusions}

In this paper we have shown that there is a phase transition in a periodic 
array of electrons each confined to a 1D ring.
The parameter $\delta$ determines if
the array will spontaneously polarize;  in 1D the transition is at
$\delta\approx 10$.  
It is easy to achieve small values of $\delta$ simply by choosing the ring
separation to be large.
Thus the ``quantum" limit where the kinetic energy
dominates is simple to obtain.  
To obtain the antiferroelectrically ordered
state we need large $\delta$.  We may write this as
$\delta=(R^2/a_0 D)\times (m^*/m)$ where $a_0$ is the Bohr radius and $m^*$
is the effective mass of the electron.
If  $\tilde R$ and $\tilde D$ are $R$ and $D$ measured in
nanometers, and $\tilde m\equiv (m^*/m)$,
then  $\delta \approx 18.9 (\tilde R^2/\tilde D)\tilde m $.  We require that
the rings do not intersect, so that $\tilde D\geq 2\tilde R$.
Thus the ability to achieve large values of $\delta$ in 
semiconductors will depend upon the value of the effective mass.  
If we set $\tilde D=2\tilde R$, then
for GaAs ($\tilde m=0.06$)  1D arrays of
rings with a radius greater than $\sim$10nm will
be polarized.  For AlAs ($\tilde m =0.4$)  the crossover radius is about
70nm.  Rings with a smaller radius will not spontaneously polarize, but
instead be isotropic

It is well known that in 1D there is no ordered state for $T>0$
for the Ising model.  However,
for small arrays over finite time intervals the system can order.
To observe this behavior we want the
characteristic energies of the system to be greater than the temperature.
For the Coulomb energy $kT<e^2/D$, which we may write as $\tilde DT< \rm
1.8\times 10^3$ where $T$ is in Kelvin.
For the kinetic energy
this means $kT<\hbar/2m^*R^2$; if we measure
$ (m/m^*) \tilde R^2 T>40$ in the same
units.  For GaAs we can choose $R$ to be about 14nm at 4K; choosing
materials with a smaller effective mass or going to lower temperature
allows us to increase the radius.

An AFE polarized ring array will scatter light at a wavelength commensurate
with the inter-ring separation, $D$.  In 1D there is a gap $\sqrt{2}\omega_0$,
which we may write as $2\sqrt{2\tilde m}  (a_0/D)^{3/2}$.   For GaAs rings
with a separation $D=1000$nm this gives $\omega\sim \rm 6.0\times 10^{10}Hz$.
The 2D arrays have a similar sized gap at zone center, but the gap vanishes
at one zone edge.  The excitation spectrum can be probed optically, but
scattering at the edge of the zone is difficult due to the constraints
imposed by conservation of energy and momentum.  Typically in such cases
Raman scattering can be used to investigate the excitations.

While we have not explicitly addressed the 2D case here, much can be 
gleaned from our results.   The 2D classical problem obviously has a finite
temperature phase transition, as shown by our Monte Carlo simulations.   The
2D quantum problem can be mapped on to the 3D XY model, which is known to order. 
We have performed simulations on the 2D case and find that it orders in
a striped phase.\cite{2Darray}

Finally these calculations assume that each ring is singly occupied.  This
might be
obtained by fabricating the rings upon a thin insulating layer covering a
gate.  By tuning the gate voltage we can bias the system so that it is
energetically favorable for an electron to tunnel to the rings.  The gate
will also serve to cutoff long distance interactions between the rings,
supporting the assumption of the nearest neighbor interactions used here.
Moreover, this letter serves to start investigation into a broad class of
problems, such as rings occupied by an optically excited exciton/hole pair
or perhaps by a small, varying number of electrons created by a random
distribution of dopants.

The topic of quantum dot arrays and their correlations has obvious and
useful analogies with solid state models of crystalline arrays of atoms.
In this paper  we wish to point out that experimentalists have at their
disposal a host of ``unnatural atoms'' analogs:  rings, quantum dot quantum
wells, quantum rice, etc.  The electrons in these nanoscale
constituents are confined to orbitals that may not have atomic analogs.
Morever, it may be possible to tune the shape of the constiutent to
optimize some desired collective property such as frustration in electric
or magnetic polarization, high susceptibility or sensitivity to optical
polarization of light.   Even more rich behavior will develop if we allow
electrons to tunnel between these nanoscale periodic structures.

\begin{acknowledgments}

The authors wish to thank Steve Girvin, Herbert Fertig, and Matthew Johnson
for several useful discussions.
This work is supported by NSF MRSEC DMR-0080054 (BR), and NSF
EPS-9720651 (KM).

\end{acknowledgments}

\appendix*
\section{Transverse Field Ising Model}
In this section we explain how one can write the projection of the hamiltonian (\ref{eq:1DTFI}) into ground and
first excited state subspace of ${\hat H}_0$ as a 1D transverse field Ising model (Eq.\ref{eq:TFIhamiltonian}).
In order to make the analysis easier we change the variables $\theta_i$ to staggered one, $(-1)^i\theta_i$.
Defining $|0\rangle$ and $|1\rangle$ as ground and first excited states of ${\hat H}_0$ and $E_0$ and $E_1$ the
corresponding eigenenergies we can write the up and down states (Fig.\ref{fig:pot}) as:
\begin{eqnarray}\label{eq:App:up-down}
|\uparrow\rangle &=& {1\over\sqrt{2}}(|0\rangle+|1\rangle) \nonumber \\
|\downarrow\rangle &=& {1\over\sqrt{2}}(|1\rangle-|0\rangle).
\end{eqnarray}
Now we define the creation and annihilation operators:
\begin{eqnarray}
c_{\uparrow}^\dagger|\uparrow\rangle &=& c_{\downarrow}^\dagger|\downarrow\rangle=0 \nonumber \\
c_{\uparrow}^\dagger|\downarrow\rangle=|\uparrow\rangle &,&
c_{\downarrow}^\dagger|\uparrow\rangle=|\downarrow\rangle ,
\end{eqnarray}
and one can show that :
\begin{equation}\label{eq:comm}
[c_{\alpha}^{},c_{\beta}^\dagger ]=\delta_{\alpha\beta}.
\end{equation}
So we can write the hamiltonian (\ref{eq:1DTFI}) as:
\begin{equation}
{\hat H}=\sum_{i=1}^{N}\sum_{\alpha,\beta=\uparrow,\downarrow}\varepsilon_{\alpha\beta}^{}c_{i\alpha}^\dagger
c_{i\beta}^{}+\sum_{\langle ij\rangle}\sum_{klmn=\uparrow,\downarrow}V_{klmn}^{ij}c_{ik}^\dagger c_{jm}^\dagger
c_{jn}^{}c_{li}^{},
\end{equation}
in which $i,j$ are spatial indices and:
\begin{eqnarray}
\varepsilon={1\over 2}\left(%
\begin{array}{cc}
  {\varepsilon_0} & {\Delta}  \\
   {\Delta} & {\varepsilon_0} \\
\end{array}%
\right)
\end{eqnarray}
where $\varepsilon_0=E_1+E_0$ and $\Delta=E_1-E_0$. To calculate matrix elements of the potential we use a
simple numerical Schrodinger equation solver to find the following quantities:
\begin{eqnarray}
\langle\uparrow|\cos\theta |\uparrow\rangle &=&\langle\downarrow|\cos\theta |\downarrow\rangle\approx 0
\nonumber \\
\langle\uparrow|\cos\theta |\downarrow\rangle &=&\langle\downarrow|\cos\theta |\uparrow\rangle\approx 0
\nonumber \\
\langle\uparrow|\sin\theta |\uparrow\rangle\approx +1 &,& \langle\downarrow|\sin\theta |\downarrow\rangle\approx
\nonumber \\
\langle\uparrow|\sin\theta |\downarrow\rangle &=&\langle\downarrow|\sin\theta |\uparrow\rangle\approx 0
\end{eqnarray}
In the above diagonal matrix elements of $\cos\theta$ are approximately zero because must of the wavefunction is
localized around $\theta=\pm\pi/2$. For the same reason the diagonal matrix elements of $\sin\theta$ are $\pm
1$. Using the above we can derive the potential in the following second quantized form:
\begin{eqnarray}
{\hat V}&=& -2\delta\epsilon^2\sum_{\langle ij\rangle} (c_{i\uparrow}^\dagger c_{j\uparrow}^\dagger
c_{j\uparrow}^{}c_{i\uparrow}^{}-c_{i\uparrow}^\dagger c_{j\downarrow}^\dagger c_{j\downarrow}^{}
c_{i\uparrow}^{}+
\nonumber\\
 &-& c_{i\downarrow}^\dagger c_{j\uparrow}^\dagger c_{j\uparrow}^{}c_{i\downarrow}^{}+c_{i\downarrow}^\dagger
c_{j\downarrow}^\dagger c_{j\downarrow}^{} c_{i\downarrow}^{}).
\end{eqnarray}
Now we can use the Holstein-Primakov transformation \cite{bosons} to construct the following SU(2) covariant
operators for each lattice point:
\begin{equation}
S^{+}=c_{\uparrow}^\dagger c_{\downarrow}^{}~,~ S^{-}=c_{\downarrow}^\dagger c_{\uparrow}^{}~ , ~S^{z}={1\over
2}(c_{\uparrow}^\dagger c_{\uparrow}^{}-c_{\downarrow}^\dagger c_{\downarrow}^{}).
\end{equation}
Using the above definitions and the fact that: $c_{\uparrow}^\dagger c_{\uparrow}+c_{\downarrow}^\dagger
c_{\downarrow}=1$ for each lattice point we arrive at the following expression for the hamiltonian
(\ref{eq:1DTFI}):
\begin{equation}
{\hat H}=-8\delta\epsilon^2\sum_{i=1}^N S_i^zS_{i+1}^z+\Delta\sum_{i=1}^N S_i^x,
\end{equation}
in which: $S^{x,y}=(S^+\pm iS^-)/2$.


\begin{thebibliography}{99}
\bibitem{limits}  The Coulomb interaction scales inversely with distance,
whereas the tunneling is exponentially suppressed.  Indeed it would be
experimentally difficult to achieve the limit where electrons {\it can}
tunnel between rings.  In the limit where electrons can tunnel from ring
to ring the dynamics are much richer ({\it unpublished}).
\end{thebibliography}
\end{document}